# Prey selection of Amur tigers in relation to the spatiotemporal overlap with prey across the Sino-Russian border

Hailong Dou[†], Haitao Yang[†], James L.D. Smith, Limin Feng, Tianming Wang, Jianping Ge

*H. Dou, H. Yang, L. Feng, T. Wang ([wangtianming@bnu.edu.cn](wangtianming@bnu.edu.cn)) and J. Ge, Ministry of Education Key Laboratory for Biodiversity Science and Engineering, Monitoring and Research Center for Amur Tiger and Amur Leopard, State Forestry and Grassland Administration & College of Life Sciences, Beijing Normal University, Beijing, 100875, China. – H. Dou, College of Life Sciences, Qufu Normal University, Qufu, China. – J. L. D. Smith, Department of Fisheries, Wildlife and Conservation Biology, University of Minnesota, St Paul, MN 55108, USA*

[†] These authors contributed equally to this work.

**Abstract**

The endangered Amur tiger (*Panthera tigris altaica*) is confined primarily to a narrow area along the border with Russia in Northeast China. Little is known about the foraging strategies of this small subpopulation in Hunchun Nature Reserve on the Chinese side of the border; at this location, the prey base and land use patterns are distinctly different from those in the larger population of the Sikhote-Alin Mountains of Russia. Using dietary analysis of scats and camera-trapping data from Hunchun Nature Reserve, we assessed spatiotemporal overlap of tigers and their prey and identified prey selection patterns to enhance understanding of the ecological requirements of tigers in Northeast China. Results indicated that wild prey constituted 94.9% of the total biomass consumed by tigers; domestic livestock represented 5.1% of the diet. Two species, wild boar (*Sus scrofa*) and sika deer (*Cervus nippon*), collectively represented 83% of the biomass consumed by tigers. Despite lower spatial overlap of tigers and wild boar compared to tigers and sika deer, tigers preferentially preyed on boar, likely facilitated by high temporal overlap in activity patterns. Tigers exhibit significant spatial overlap with sika deer, likely favoring a high level of tiger predation on this large-sized ungulate. However, tigers did not preferred roe deer (*Capreolus pygargus*) and showed a low spatial overlap with roe deer. Overall, our results suggest that tiger prey selection is determined by prey body size and also overlap in tiger and prey use of time or space. Also, we suggest that strategies designed to minimize livestock forays into forested lands may be important for decreasing the livestock depredation by tigers. This study offers a framework to simultaneously integrate food habit analysis with the distribution of predators and prey through time and space to provide a comprehensive understanding of foraging strategies of large carnivores.

**Introduction**

The tiger (*Panthera tigris*), a charismatic species and the largest of the extant cats in the world, has lost 93% its historical range during the past century (Dinerstein et al. 2007). Habitat loss, poaching, and widespread wild prey depletion have been the principle contributors to the tigers decline over the last several decades. Tigers now persist in increasingly isolated and often degraded habitats, and are on the brink of local extinction in many locations (Gopal et al. 2010, Walston et al. 2010, Wang et al. 2016). Tigers generally perform better and reach higher densities where the prey density of medium to large wild ungulates is high (Karanth et al. 2004, Miquelle et al. 2010). Prey selection by large felids plays a fundamental role in defining their geographical distribution, dispersal, habitat selection, and social structure (Karanth et al. 2004, Petrunenko et al. 2016, Sunquist and Sunquist 1989). Continued decreases in the density of key prey species may currently be one of the major threats to tiger persistence in many areas (Hebblewhite et al. 2012, Karanth and Stith 1999, Sandom et al. 2017, Wang et al. 2016).

Knowledge of both tiger diet and the abundance of its primary prey are critical to recovering small and threatened populations (Biswas and Sankar 2002, Khorozyan et al. 2015). It is the case with the endangered Amur (Siberian) tiger (*P. t. altaica*), which occurs on the most northern edge of the tiger's range. Currently, fewer than 600 individuals are estimated to remain in two isolated subpopulations confined to the Sikhote-Alin Mountains of Russia and the Changbai Mountains along the China-Russia border (Miquelle et al. 2010).

Since the late 1990s, the Changbai Mountains trans-boundary subpopulation has gradually increased to approximately 40 individuals and is extending its distribution into China (Feng et al. 2017, Wang et al. 2015, Wang et al. 2014). However, growth of this trans-boundary subpopulation may be limited by conflict with humans in the form of cattle depredations (Soh et al. 2014). Perhaps more importantly, Wang et al. (2016) reported that cattle grazing degrades tiger habitat by negatively influencing the abundance and distribution of major ungulate prey. In particular, cattle (*Bos taurus*) have displaced the sika deer (*Cervus nippon*) and have become a major constraint to reestablishment of Amur tigers in Northeast China (Wang et al. 2018) because the distribution and territory of tigers are closely associated with those of their principal prey (Karanth et al. 2004, Miquelle et al. 2010). Female Amur tigers require 4-20 times more land (*ca.* 400 km$^2$ home range) than that of any other Asian tiger populations (Hernandez-Blanco et al. 2015, Miquelle et al. 2010). To conserve a minimum viable tiger population in this region requires securing a much larger prey and land base (Wang et al. 2016, Wang et al. 2018). An analysis of tiger diet is critical to the recovery of this population.

Most information on prey selection of Amur tiger comes from studies carried out in southwest Primorye and the Sikhote-Alin Mountains of the Russian Far East. In those areas, despite tigers having a relatively broad dietary range, consuming approximately 15 different prey species, their diet is uniformly dominated by red deer (*Cervus elaphus*), wild boar (*Sus scrofa*), sika deer and roe deer (*Capreolus pygargus*), which collectively constitute 83-90% of biomass consumed, with other wild prey and domestic species contributing little to tiger diet (Kerley et al. 2015, Miquelle et al. 2010, Miquelle et al. 1996, Sugimoto et al. 2016). The abundance and vulnerability of preferred food resources (red deer and wild boar) in the landscape are proposed to be the key driving force in determining the habitat use and home range of the Russian Sikhote-Alin tiger population (Petrunenko et al. 2016).

Classical diet analysis approaches, such as kill composition and scat analysis, can provide a valid assessment of carnivore predation (Karanth and Sunquist 1995, Kerley et al. 2015). However, these methods have inherent limitations in their ability to simultaneously address the spatiotemporal contact of predator and prey, which may lead to an incomplete understanding of the foraging strategies of solitary predators. Camera traps can document activity patterns and space use from multiple locations within a short time. Camera traps have recently been widely used to augment food habit studies of many carnivores, including jaguar (Weckel et al. 2006), dhole (Kamler et al. 2012) and leopard (Braczkowski et al. 2012, Henschel et al. 2011). However, where or when Amur tigers and their prey co-occur and how tiger-prey interactions influence tiger predation are poorly understood and therefore a research priority for management and conservation of Amur tigers.

Here, we investigate, for the first time, the foraging strategies of Amur tigers in Sino-Russian international border zones using scat analysis combined with camera trapping data. Our objectives are to determine: 1) which species constitute the dominant winter prey of tigers, 2) which prey species are selectively preyed upon by tigers, and 3) what are the similarities and differences between tiger and prey activity and temporal use patterns. We then compare the diet in Hunchun Reserve with other diet studies of Amur tigers.

**Methods**

**Study area**

The Hunchun Nature Reserve (HNR) established in 2001 is situated in eastern Jilin Province, China, with a total area 1087 km$^2$ (Fig. 1). The reserve borders the Land of Leopard National Park in Southwest Primorsky Krai, Russia, forming a trans-boundary conservation landscape (Wang et al. 2016, Wang et al. 2016). The terrain of HNR is hilly to mountainous, with elevation ranging from 5 to 937 m above sea level. The climate is temperate continental monsoon with mean annual temperature ranging from 3.90 to 5.65 °C, a frost-free period of 120-126 days per year, and mean annual precipitation of 618 mm during 1990-2010. Vegetation is a mixed Korean pine (*Pinus koraiensis*)-deciduous forest dominated by Korean pine, Mongolian oak (*Quercus mongolica*), Manchurian walnut (*Juglans mandshurica*), Manchurian ash (*Fraxinus mandshurica*), and maple (*Acer* spp.) and birch (*Betula* spp.). More than 80% of forests have been logged, and nearly 95% of low-elevation forests have been converted into secondary deciduous forests over the past 5 decades (Li et al. 2009, Xiao et al. 2014). Since 1998, logging of natural forests has been halted. More than 14,000 people live in 29 villages within the reserve, and the average people density is 12 people/km$^2$ (Xiao et al. 2016). The main economic activity within HNR is free-range cattle grazing; other human activities include the collection of edible ferns, ginseng farms, and frog farming (Wang et al. 2016).

In addition to the Amur tiger, the study area included Amur leopard (*Panthera pardus orientalis*), Eurasian lynx (*Lynx lynx*), black bear (*Ursus thibetanus*) and brown bear (*Ursus arctos*), and potential prey species, such as the sika deer, Siberian roe deer, wild boar and musk deer (*Moschus moschiferus*) (Miquelle et al. 2010, Tian et al. 2011, Xiao et al. 2014). Cattle and dogs (*Canis lupus familiaris*) were also a common component of the tiger diet.

**Scat analysis**

Tiger diet was evaluated using scats (i.e., faeces) that were opportunistically collected by walking a network of small trails, ridgelines, stream beds used by ungulates and tigers as well as forest roads. These routes were patrolled systematically by trained field staff and researchers in HNR in search of scats from November 2014 to April 2015 (Fig.1) (Dou et al. 2016) (Fig. 1). We identified tiger scats based on mitochondrial DNA analysis (Dou et al. 2016). After species identification, the tiger scat was thoroughly washed several times over a 0.7 mm screen until prey remains, such as hair and other undigested body parts, were separated from the scat. From each scat, a predefined minimum of 20 hairs were sampled and hairs were identified to species by examining the general appearance (e.g., colour, width, length and tortuosity), structure patterns of the cuticle and medulla, and cross sections under a microscope and comparing these to a reference collection of hairs from a standard prey hair manual for Amur tigers (Rozhnov et al. 2011).

We used percent occurrence (i.e., percentage of scats containing a particular food item) and percent biomass consumed to quantify the contribution of each species to the tiger diet. For scats containing two prey items, each scat was counted as 0.5 prior to calculating the percent occurrence and percent biomass (Karanth and Sunquist 1995). We used a nonlinear (asymptotic) model (biomass consumed per collectable scat or predator weight $=0.033-0.025\exp^{-4.284(\text{prey weight/predator weight})}$) developed by Chakrabarti et al. (2016) to calculate prey biomass consumed from scats. The mean live weights of tiger and different prey species were obtained from previous studies (Bromley and Kucherenko 1983, Danilkin 1999, Miller et al. 2014) (the prey species that weighed < 2 kg were excluded). Finally, we calculated the percent biomass contribution using the above equation (biomass of each prey type consumed/total biomass consumed × 100) (Ramesh et al. 2009). We estimated 95% bootstrapped confidence intervals of percent occurrence and percent biomass contributions based on 10,000 replicates with replacement using percentile method in R package *boot*.

**Camera trap data collection**

We employed camera trap field data to assess the abundance, activity patterns and distribution of Amur tigers and potential prey species within the study area. This study, conducted from November 2014 to April 2015, was part of a long-term Tiger Leopard Observation Network (TLON) project that employed camera trap stations in Hunchun Nature Reserve and its surrounding area (Wang et al. 2016). A total of 104 camera trap stations were used in this study (Fig. 1). We used 3.6 × 3.6 km$^2$ grids to guide camera trap placement throughout the study area. Within the sampling grids, we maximized the detection probability by placing cameras at sites where tigers, leopards, and their prey are likely to travel (e.g., along ridges, valley bottoms, trails, forest roads and near scent-marked trees). We deployed cameras along forest roads (n =48 sites) and game trails (n =56 sites). We excluded farmland and villages. The cameras (LTL 6210M, Shenzhen, China) were fastened to trees approximately 40-80 cm above the ground and were programmed to take photographs 24 hours/day with a 1-minute interval between consecutive events. We report the number of detections and number of trap stations detected for each species. To avoid inflated counts caused by repeated detections of the same event, only one record of a species at a trap site was recorded per 0.5 hour.

**Abundance of prey species**

Because the number of detections of each species is dependent on a unique set of ecological factors we did not use a relative abundance index to estimate prey selectivity (Sollmann et al. 2013). Instead the abundance of three major prey species (i.e., wild boar, roe deer and sika deer) at each camera station was estimated using N-mixture models (Royle 2004) with camera days (total days each camera was in operation) as predictors of detection. We also allowed for time varying detection probabilities within different occasions. N-mixture models assume that all within-site variation in counts is attributable to detection probability and no false-positives occur (i.e., detecting individuals more than once) (Kery and Royle 2015). This approach is suitable when it is impossible to distinguish individuals of the species and is a reasonable surrogate for abundance (Kery and Royle 2015). Recent studies confirm that N-mixture models can provide reliable estimates of relative abundance despite the challenge of ensuring complete population closure within a sampling occasion (Barker et al. 2017, Denes et al. 2015). For each camera site, we used a 2-week periods as the temporal sampling unit (i.e., survey occasion) and counted the number of individuals in each 'event', an independent 15-second video (we considered 'events' occurring > 30 minutes apart as independent), and then calculated the accumulated individuals within each 2-week occasion. All models used Poisson distribution and were conducted in the R package *unmarked* (Fiske and Chandler 2011).

**Prey selectivity**

Prey selection, or feeding preferences, of tigers was estimated for each species by comparing the observed proportion of prey items in scats (i.e., utilization) with the expected proportion of potential prey in the environment (i.e., availability). We restricted our analyses of preference to three species: wild boar, roe deer and sika deer because we lacked data on the relative abundance of other prey species. Based on the utilization and availability of each prey species, Jacobs's index (Jacobs 1974) was calculated: $D = (r_i - p_i)/(r_i + p_i - 2r_ip_i)$, where $r_i$ is the percent occurrence of prey item *i* obtained by the scat analysis, and $p_i$ is the proportional abundance of prey item *i* obtained by the N-mixture model. The index values range from - 1 (strongly avoided) to + 1 (strongly preferred).

**Overlap in space use**

We assessed the potential spatial associations between tigers and prey species using two different approaches. First, following Ramesh et al. (2012), we calculated Pianka's index (*O*) between tigers and major prey (Pianka 1973) using the presence of each species at each camera station in the *spaa* package in the R software (Zhang et al. 2013). Next, we used a single-species single-season occupancy model to evaluate the habitat use of tigers while accounting for the imperfect detection (MacKenzie 2006). Given there were multiple camera trap stations within each tiger 's home range, it is habitat use, rather than occupancy that we are modeling. We assumed animals move randomly between the fine-scale sampling sites, which relaxed the assumption of geographical closure typically required for occupancy models. We defined 2-week periods as temporal replicates and constructed detection histories of tigers and for each camera station over 13 sampling occasions. We considered the relative abundance (number of detections acquired/100 trap days) of three major prey species as predictors of tiger occupancy, the trail type (forest road or game trail at each camera location) and camera days (effort) as predictors of detection. Occupancy model also was implemented in the *unmarked* package. To assess model goodness-of-fit, we used 1000 parametric bootstrap samples on a chi-square test that is appropriate for binary data.

**Activity pattern and temporal overlap**

We defined dawn and dusk time periods as 1 h prior to and 1 h post sunrise and sunset, respectively (Farris et al. 2015). Species primarily active during dawn and dusk are referred to as crepuscular. We defined diurnal time period as between dawn and dusk, whereas nocturnal time period was between dusk and dawn. For this, program Moonrise 3.5 (http://www.rocketdownload.com/program/moonrise-424593.html) was used to determine the daily times of sunrise and sunset during the study period. Consequently, the activity times of each independent event per species were classified into three categories: crepuscular (05:01-7:00 and 15:44-17:43 h), diurnal (7:01-15:43 h) and nocturnal (17:44-05:00 h). Then, the activity patterns of each species were classified into the following categories: crepuscular (approximately 50% of detections during the crepuscular phase), diurnal (< 10% of detections in the night), nocturnal (> 90% of detections in the night), mostly diurnal (10% ~ 30% of detections in the night), mostly nocturnal (70% ~ 90% of detections in the night) and cathemeral (the rest of the detections) (Jiménez et al. 2010). We used kernel density estimation to quantify the activity density of tigers and their prey types. Then, we used the R package 'overlap' to estimate the overlapping of activity patterns between them (Ridout and Linkie 2009). To assess whether tigers and prey differed in their diel activity patterns, we tested the percent photographic capture for each hour using a Spearman's rank correlation coefficient test (Spearman's rho).

Results

**Species composition of the tiger diet**

A total of 148 scats were collected in our analyses. Among them, 118 tiger scats and 3 leopard scats were confirmed by DNA. Three tiger scats were removed from further analysis because they contained grass, leaves or unidentifiable dietary remains. The remaining 27 scats were unidentifiable using DNA and discarded from analysis. At least 9 tigers were identified from the scats used, which was confirmed by our previous DNA analysis (Dou et al. 2016). Six prey species were identified in tiger faeces (Table 1). Six scats (5%) contained two prey items. Wild ungulates constituted 93.82% of the total biomass consumed, followed by domestic species (5.15%) and mustelids (1.03%). Wild boar and sika deer were two most dominant prey species in terms of biomass consumed (83%) in tiger scats (Table 1).

**Prey abundance and selection**

From November 2014 to April 2015, a total of 1288 independent photographs of 19 potential prey species were obtained over 17,048 trap days (Table 2). We also obtained 131 photos of 11 tigers (5 males and 6 females). Tigers triggered 40% of all camera stations. Based on N-mixture model results, relative abundance (±SE) of wild boar, roe deer, and sika deer were 1.90 (±0.50), 3.30 (±0.70), and 2.18 (± 0.28), respectively, in the winter of 2014-2015. The three ungulates were photographed at 45%, 52% and 43% of the stations, respectively (Table 2 and Fig. 2). The detection probability of three ungulates varied among 2-week time periods and was strongly influenced by sampling effort (camera days, for wild boar $\beta$ = 0.53, SE = 0.25, $p$ = 0.03, for roe deer $\beta$ = 0.66, SE = 0.23, $p$ <0.01, for sika deer $\beta$ = 0.68, SE = 0.22, $p$ <0.01). The percent occurrence of prey species consumed by tigers did not reflect the abundance available; tigers showed a notable preference for wild boar ($D$ = 0.58, 95% CI: 0.38-0.64), appeared to use sika deer similarly to their availability ($D$ = -0.02, 95% CI: -0.29-0.11) but roe deer was not preferred ($D$ = -0.65, 95% CI: -0.81- -0.53) (Fig. 3).

**Spatiotemporal overlap between the tiger and its main prey species**

Tigers spatially overlapped with sika deer ($O$ = 0.35) to a greater extent than with wild boar ($O$ = 0.16) and roe deer ($O$ = 0.18) (Table 3 and Fig. 2). Our occupancy model confirmed that tiger habitat use was significantly positively correlated with sika deer spatially but were considerably negatively correlated with roe deer (Table 4). Tigers were estimated to occur across 56% (95% CI: 44-68%) of the camera trap stations. Our chi-square statistic indicated no evidence for lack of model fit ($p$ = 0.185).

Tigers showed a strong nocturnal and crepuscular pattern (78.8% of detections between sunset and sunrise) but exhibited peaks of activity around dawn and dusk (Table 5; Fig. 4). Wild boar exhibited similar activity patterns but showed peaks around dusk. The activity patterns of sika deer and roe deer were mostly diurnal, but roe deer exhibited peaks of activity around twilight, tending towards being crepuscular (Table 5; Fig. 4). Temporal activity of tigers was significantly correlated with wild boar temporally (Spearman's rho = 0.22, $p$ < 0.01) with temporal overlap $\Delta$ = 0.76 (95% CI: 0.67-0.85), but were significantly not in sync with sika deer (Spearman's rho = -0.42, $p$ < 0.01) and the temporal overlap was low ($\Delta$ = 0.69, 95% CI: 0.61-0.78) (Table 3; Fig. 4). Activity of tiger and roe deer was not correlated significantly (Spearman's rho = 0.02, $p$ < 0.01) with temporal overlap $\Delta$ = 0.74

(95% CI: 0.65-0.83).

**Discussion**

**Species composition of tiger diet**

This study reports the critical prey resources for the recovery of this small population of tigers in China. Our camera traps recorded 11 tigers during the study period. All scats used in this study came from at least 9 tiger individuals, indicating that majority (82%) of HNR tigers detected using camera traps contributed to the scat samples but with high heterogeneity in the number of scats of each tiger (mean =9.6, SD =9.2). At the population level, tigers are known to be selective predators. Our results are in accordance with previous findings regarding the diet of tigers across their range, which indicate that medium to large wild ungulates formed the majority of the tiger diet (Karanth and Sunquist 1995, Sugimoto et al. 2016). Wild boar, sika deer and roe deer contribute up to > 90% of the total biomass consumed (Table 1), illustrating that they are currently key prey for this small tiger population across the Sino-Russia border. However, our results differed from those of Miquelle et al. (2010) in the Russian Far East in that red deer (200 kg) were another very important prey item.

To more deeply explore the Amur tiger dietary requirements and differences in winter, we compared our results with those from a study by Kerley et al. (2015) implemented at three sites in Russia. We had only 6 prey species compared to 9 or 10 at 3 other sites (Table S1) (Kerley et al. 2015). The lower diversity at our site during winter reflects the absence of several prey species found elsewhere and also may be a consequence of our smaller sample size and lower study duration. Sika deer have largely replaced red deer as the most common cervid in HNR, China, and adjacent SW, Russia. Kerley et al.'s other two sites were areas where the prey base included red deer that are not found in HNR and SW, Russia (Table S1). As reported by Griffiths (1975), tigers are also 'energy maximizers' and in our study with the two largest ungulates, wild boar and sika deer, contributed > 80% of the total biomass consumed by tigers (Table 1). However, compared with wild boar, sika deer were consumed to a lesser extent by tigers, even though they had the highest abundance among the ungulates. This result is partially attributable to the difference in this two prey's vulnerability and other factors (see below). Small prey species (e.g., badgers) were occasionally preyed upon, reflecting the avoidance of smaller prey by tigers (Karanth and Sunquist 1995, Sugimoto et al. 2016).

Cattle were identified in the tiger diet in winter, but represents a relatively low contribution to their diet (2.88% of the total biomass consumed). The overall biomass contribution of cattle may be underestimated, because almost all cattle were brought back to the villages from the forest in winter. Predation on cattle has been associated with easy access linked with poor livestock husbandry practices (Wang et al. 2016); more than 30% of the HNR and its surrounding area is grazed by unattended domestic livestock at an average stocking rate of 8-12 cattle /km$^2$ in summer. Tigers frequently prey on cattle, suggesting that cattle have become a regular food source for this border population. However, this behaviour has resulted in substantial human-tiger conflict, as reported by Wang et al. (2016) and Soh et al. (2014), who reported that more than 370 cattle depredations occurred between 2008 and 2014. A similar pattern was found in a different national park in India, where livestock accounted for 6 to 12% of the tiger diet despite the park's high wild prey densities (Bagchi et al. 2003, Biswas and Sankar 2002, Reddy et al. 2004).

**Prey abundance and selection**

The body size, availability and vulnerability of prey are the primary factors determining prey selection. A recent meta-analysis (Hayward et al. 2012) suggests that species weighing between 60 and 250 kg are preferred by tigers, and in our study the two large prey, wild boar (103 kg) and sika deer (95 kg) were the preferred prey. The wild boar is the largest ungulate preferred by Amur tigers (Fig. 3), supporting the findings of Hayward et al. (2012); they report that the wild boar is one of the species most preferred by tigers based on selectivity index scores from 3187 kills or scats from 32 prey species. Our results are also congruent with those of earlier studies from reserves in Russia (Kerley and Borisenko 2007, Kerley et al. 2015, Sugimoto et al. 2016). The strong preference for wild boar may reflect its behaviour, especially when in groups, of noisily foraging, head down in the leaf litter making it much easier to approach. Groups of wild boar are likely slower when attempting to escape when the ground is covered with snow in winter (Miquelle et al. 2010). Easy to locate and prey wild boar may explain their preferred status by tigers. Previous study also showed that individual vigilance decreased with increasing group size when wild board were feeding (Quenette and Gerard 1992).

Sika deer are the second most dominant prey species of tigers at our study site, and tigers showed a neutral preference toward this species. Sika deer may be killed less frequently than expected based on their relative abundance within three ungulate prey. However, we believe high abundance of sika deer in our study area makes them a key prey resource. Furthermore, large sized deer generally represent a dominant proportion of a tiger's prey biomass requirement in most parts of its range (Biswas and Sankar 2002, Hayward et al. 2012). Our results differed from those of Kerley et al. (2015) and Sugimoto et al. (2016) in adjoining SW Russia, where tigers selected against sika deer ($D < -0.65$). Preference differences may reflect differences in species specific hunting success among sites. For example, snow depth and terrain differences may explain the differences in preference. Further research is needed to better understand how the probability of encounter and hunting success contribute to prey preference.

The relative abundance index has become a common approach to measure prey available in the study of animal diet (e.g. Weckel, Giuliano & Silver, 2006 and Henschel *et al.*, 2011). But this index is highly biased by a suite of factors (e.g., species ecology, imperfect detection and study design) (Sollmann et al. 2013), we used prey abundance estimated from N-mixture models, which accounted for imperfect detection of individuals, to determine the relative availability of major prey.

**Spatiotemporal overlap between the tiger and its main prey species**

As diet largely reflects foraging strategies of the big cats, spatiotemporal overlap between predator and prey may affect the composition of carnivore diet (Kronfeld-Schor and Dayan 2003, Weckel et al. 2006). Our research supports this assertion. We discovered that tigers are mainly nocturnal and crepuscular in our study area, which was similar to the activity reported for tigers in Nepal and India (Carter et al. 2012, Karanth et al. 2017) but contrasted with that of tigers in Malaysia (Rayan and Linkie 2015), where tigers showed a strong diurnal pattern. These differences were partially reflected in tiger-prey temporal interaction across the sites. Apparently, the activity of tigers maximizes their encounters with wild boar, the ungulate species that was most preferred, despite their relatively low abundance and obvious low spatial overlap with tigers. This result suggests that synchronized activity between tigers and wild boar likely facilitates a high level of tiger predation on this prey species which may represent an optimization of foraging time to maximize energetic gain. Contrary to our prediction, however, the activity of tigers was considerably different from that of the 2 deer species they selectively consumed (i.e., roe and sika deer). Tigers exhibited a similar space use pattern to that of sika deer. This indicates that Amur tigers use habitat where sika deer are densely populated, as also shown by Wang et al. (2016) in our study area. Conversely, the spatiotemporal partitioning reduced the possibility for chance encounters between tigers and the mostly diurnal roe deer, revealing that tigers may opportunistically prey on roe deer. In terms of energy rewards, roe deer are poor-quality prey for tigers because of the high energetic costs involved in capturing this small, wary and active species (Miller et al. 2014). Overall, the temporal or spatial synchronization of two large ungulates (sika deer and wild boar) with tigers increased their encounter likelihood at each camera-trap station, resulting in tigers preferentially hunting them. Also, we would like to point out that camera placement decisions and biological interactions (e.g. landscape of fear and interaction between ungulates) may influence the spatial overlap between tiger and prey, which requires further study.

It has been demonstrated that carnivore habitat use is focused on those areas where prey are more vulnerable and/or where prey are more abundant (Balme et al. 2007, Hopcraft et al. 2005, Petrunenko et al. 2016). In Sikhote-Alin Biosphere Zapovednik of the Russian Far East, Petrunenko et al. (2016) discovered that tiger habitat use and home range establishment are affected by the abundance and vulnerability of red deer and wild boar in the landscape but are not significantly affected by those of sika deer or composite maps where all three prey species occur together. Our results might suggest that tigers use habitat within their home range where sika deer and wild boar are densely populated due to red deer's absence in our region.

**Limitations of the study and ways forward**

Like many other studies of the food habits of tigers based on scat analysis, our results have several limitations. In addition to the use of a small number of scat samples, this study did not consider selectivity in terms of the age and sex of prey. By examining kill sites, Miller et al. (2013) reported that Amur tigers present selectivity based on the sex and age of ungulates. For example, more than 67% of wild boar killed by tigers are subadults and piglets. Hence, the relative contribution of wild boars to the diet of tigers will decline when the body size of kills is taken into account. Given that the N-mixture models estimated the relative abundance of prey species (Barker et al. 2017), the lack of an independent assessment of animal densities in the environment was another substantial limitation in this study. The combination of multiple field methods, including camera trapping, faecal counts and transect line surveys, to enhance abundance estimates should be pursued.

It is worth noting that the primary prey also exhibited relatively low spatial overlap with tigers in winter ($O$ = 0.16-0.35, Table 3) compared to the summer season (unpublished data). We speculated that snow is the main reason for this low spatial overlap because ungulates concentrated in valleys and their activity levels were reduced in the snowy winters (Miquelle et al. 1996). The activity ranges of large predators, such as Amur tigers, are consistent both seasonally and over multiple years (Hojnowski et al. 2012). A further study examining the diet and spatial and temporal habitat use in seasons other than winter will help us better understand resource use by tigers. Finally, competition with sympatric leopards and anthropogenic disturbances (human presence, cattle grazing, poaching, etc.) are also considered to be factors affecting prey abundance and the spatial use of tigers, and the mechanisms remain unclear. Therefore, future studies should seek to understand the impact of people and leopards on the food habits of tigers.

**Conservation implications**

Overall, prey selection by tigers is not just dependent on the body size of ungulates but apparently also on their behavioural flexibility in exploiting prey. The use of camera traps has greatly increased our ability to examine tiger-prey temporal and spatial relationships. The approach we applied can be used as a framework to simultaneously integrate food habit analysis with the distribution of predators and prey through time and space to enhance the comprehensive understanding of tiger foraging strategies. HNR supports a very low population of tigers (0.33-0.40/100 km$^2$) (Xiao et al. 2016) presumably because of low prey densities resulting from habitat degradation. However, increased cattle grazing now has decreased the abundance of ungulate species and have become a major hurdle to the recovery of Amur tigers in China (Wang et al. 2016). We suggest that conservation concerns should be focused on implementing practices to gradually reduce cattle grazing levels and extent and increase the size of sika deer and wild boar populations in the park.

*Acknowledgements* – We thank the State Forestry Administration, the Jilin Province Forestry Bureau, and the Forestry Industry Bureau of Heilongjiang Province for kindly providing research permits and facilitating fieldwork. We also thank our field survey staff, Tonggang Chen, Shuyun Peng, Wenhong Xiao, Xiaodan Zhao and Yanchao Cheng, for collecting the scat samples. We are grateful to Raj Kumar Baniya for advice on scat analysis protocols.

*Funding* – This work was supported by grants from the National Natural Science Foundation of China (31470566, 31210103911 and 31270567), the National Key Research and Development Program (2016YFC0500106), and the Open Project of MOE Key Laboratory for Biodiversity Science and Ecological Engineering (K201703). J.L.D. Smith's contribution to this research was supported by the USDA National Institute of Food and Agriculture.

# References

Bagchi, S., et al. 2003. Prey abundance and prey selection by tigers (*Panthera tigris*) in a semi-arid, dry deciduous forest in western India. - J. Zool. 260: 285-290.

Balme, G., et al. 2007. Feeding habitat selection by hunting leopards *Panthera pardus* in a woodland savanna: prey catchability versus abundance. - Anim. Behav. 74: 589-598.

Barker, R. J., et al. 2017. On the reliability of N-mixture models for count data. - Biometrics: DOI: 10.1111/biom.12734.

Biswas, S. and Sankar, K. 2002. Prey abundance and food habit of tigers (*Panthera tigris tigris*) in Pench National Park, Madhya Pradesh, India. - J. Zool. 256: 411-420.

Braczkowski, A., et al. 2012. Diet of leopards in the southern Cape, South Africa. - Afr. J. Ecol. 50: 377-380.

Bromley, G. F. and Kucherenko, S. P. 1983. Ungulates of the southern Far East USSR. - Nauka (Science) Press.

Carter, N. H., et al. 2012. Coexistence between wildlife and humans at fine spatial scales. - Proc. Natl. Acad. Sci. USA 109: 15360-15365.

Chakrabarti, S., et al. 2016. Adding constraints to predation through allometric relation of scats to consumption. - J. Anim. Ecol. 85: 660-670.

Danilkin, A. A. 1999. Mammals of Russia and adjacent regions: deer (*Cervidae*). - GEOS.

Denes, F. V., et al. 2015. Estimating abundance of unmarked animal populations: accounting for imperfect detection and other sources of zero inflation. - Methods Ecol. Evol. 6: 543-556.

Dinerstein, E., et al. 2007. The fate of wild tigers. - Bioscience 57: 508-514.

Dou, H. L., et al. 2016. Estimating the population size and genetic diversity of Amur tigers in Northeast China. - Plos One 11: e0154254.

Farris, Z. J., et al. 2015. When carnivores roam: temporal patterns and overlap among Madagascar's native and exotic carnivores. - J. Zool. 296: 45-57.

Feng, L. M., et al. 2017. Collaboration brings hope for the last Amur leopards. - Cat News 65: 20.

Fiske, I. J. and Chandler, R. B. 2011. Unmarked: An R package for fitting hierarchical models of wildlife occurrence and abundance. - J. Stat. Softw. 43: 1-23.

Gopal, R., et al. 2010. Evaluating the status of the Endangered tiger *Panthera tigris* and its prey in Panna Tiger Reserve, Madhya Pradesh, India. - Oryx 44: 383-389.

Griffiths, D. 1975. Prey availability and the food of predators. - Ecology 56: 1209-1214.

Hayward, M. W., et al. 2012. Prey preferences of the tiger *Panthera tigris*. - J. Zool. 286: 221-231.

Hebblewhite, M., et al. 2012. Is there a future for Amur tigers in a restored tiger conservation landscape in Northeast China? - Anim. Conserv. 15: 579-592.

Henschel, P., et al. 2011. Leopard prey choice in the Congo Basin rainforest suggests exploitative competition with human bushmeat hunters. - J. Zool. 285: 11-20.

Hernandez-Blanco, J. A., et al. 2015. Social structure and space use of Amur tigers *(Panthera tigris altaica)* in Southern Russian Far East based on GPS telemetry data. - Integr. Zool. 10: 365-375.

Hojnowski, C. E., et al. 2012. Why do Amur tigers maintain exclusive home ranges? Relating ungulate seasonal movements to tiger spatial organization in the Russian Far East. - J. Zool. 287: 276-282.

Hopcraft, J. G. C., et al. 2005. Planning for success: Serengeti lions seek prey accessibility rather than abundance. - J. Anim. Ecol. 74: 559-566.

Jacobs, J. 1974. Quantitative measurement of food selection. - Oecologia 14: 413-417.

Jiménez, C. F., et al. 2010. Camera trap survey of medium and large mammals in a montane rainforest of northern Peru. - Rev. Peru. Biol. 17: 191-196.

Kamler, J. F., et al. 2012. The diet, prey selection, and activity of dholes (*Cuon alpinus*) in northern Laos. - J.

Mammal. 93: 627-633.

Karanth, K. U., et al. 2004. Tigers and their prey: Predicting carnivore densities from prey abundance. - Proc. Natl. Acad. Sci. USA 101: 4854-4858.

Karanth, K. U., et al. 2017. Spatio-temporal interactions facilitate large carnivore sympatry across a resource gradient. - P. Roy. Soc. B-Biol. Sci. 284.

Karanth, K. U. and Stith, B. M. 1999. Prey depletion as a critical determinant of tiger population viability. - In: Seidensticker, J., et al. (eds.), Riding the Tiger. Tiger Conservation in Human-dominated Landscapes. Cambridge University Press, pp. 110-113.

Karanth, K. U. and Sunquist, M. E. 1995. Prey selection by tiger, leopard and dhole in tropical forests. - J. Anim. Ecol. 64: 439-450.

Kerley, L. L. and Borisenko, M. I. 2007. Using scat detection dogs to collect Amur leopard and tiger scats for comparative analysis. A final report to the Wildlife Conservation Society.

Kerley, L. L., et al. 2015. A comparison of food habits and prey preference of Amur tiger (*Panthera tigris altaica*) at three sites in the Russian Far East. - Integr. Zool. 10: 354-364.

Kery, M. and Royle, J. A. 2015. Applied Hierarchical Modeling in Ecology: Analysis of distribution, abundance and species richness in R and BUGS: Volume 1: Prelude and Static Models. - Academic Press.

Khorozyan, I., et al. 2015. Big cats kill more livestock when wild prey reaches a minimum threshold. - Biol. Conserv. 192: 268-275.

Kronfeld-Schor, N. and Dayan, T. 2003. Partitioning of time as an ecological resource. - Annu. Rev. Ecol., Evol. Syst. 34: 153-181.

Li, Z. W., et al. 2009. Land use pattern and its dynamic changes in Amur tiger distribution region. - Chin. J. Appl. Ecol. 20: 713-724.

MacKenzie, D. I. 2006. Occupancy estimation and modeling: inferring patterns and dynamics of species occurrence. - Academic Press.

Miller, C. S., et al. 2013. Estimating Amur tiger (*Panthera tigris altaica*) kill rates and potential consumption rates using global positioning system collars. - J. Mammal. 94: 845-855.

Miller, C. S., et al. 2014. Amur tiger (*Panthera tigris altaica*) energetic requirements: Implications for conserving wild tigers. - Biol. Conserv. 170: 120-129.

Miquelle, D. G., et al. 2010. The Amur tiger: a case study of living on the edge. - In: Macdonald, D. W. and Loveridge, A. J. (eds.), Biology and conservation of wild felids. Oxford University Press, pp. 325-339.

Miquelle, D. G., et al. 1996. Food habits of Amur tigers in Sikhote-Alin Zapovednik and the Russian Far East, and implications for conservation. - J. Wildl. Res. 1: 138-147.

Petrunenko, Y. K., et al. 2016. Spatial variation in the density and vulnerability of preferred prey in the landscape shape patterns of Amur tiger habitat use. - Oikos 125: 66-75.

Pianka, E. R. 1973. The structure of lizard communities. - Annu. Rev. Ecol. Syst. 41: 53-74.

Quenette, P. Y. and Gerard, J. F. 1992. From Individual to Collective Vigilance in Wild Boar (*Sus Scrofa*). - Can J Zool 70: 1632-1635.

Ramesh, T., et al. 2012. Spatio-temporal partitioning among large carnivores in relation to major prey species in Western Ghats. - J. Zool. 287: 269-275.

Ramesh, T., et al. 2009. Food habits and prey selection of tiger and leopard in Mudumalai Tiger Reserve, Tamil Nadu, India. - J. Sci. Trans. Environ. Technov. 2: 170-181.

Rayan, D. M. and Linkie, M. 2015. Conserving tigers in Malaysia: A science-driven approach for eliciting conservation policy change. - Biol. Conserv. 184: 18-26.

Reddy, H. S., et al. 2004. Prey selection by the Indian tiger (*Panthera tigris tigris*) in Nagarjunasagar Srisailam

Tiger Reserve, India. - Mamm. Biol. 69: 384-391.

Ridout, M. S. and Linkie, M. 2009. Estimating overlap of daily activity patterns from camera trap data. - J Agr Biol Envir St 14: 322-337.

Royle, J. A. 2004. N-mixture models for estimating population size from spatially replicated counts. - Biometrics 60: 108-115.

Rozhnov, V. V., et al. 2011. A guide to deer species in the diet of Amur tiger(microstructure of deer species guard hairs found in Amur tiger excrement). - Scientific Press.

Sandom, C. J., et al. 2017. Learning from the past to prepare for the future: Felids face continued threat from declining prey richness. - Ecography: DOI: 10.1111/ecog.03303.

Soh, Y. H., et al. 2014. Spatial correlates of livestock depredation by Amur tigers in Hunchun, China: Relevance of prey density and implications for protected area management. - Biol. Conserv. 169: 117-127.

Sollmann, R., et al. 2013. Risky business or simple solution - Relative abundance indices from camera-trapping. - Biol. Conserv. 159: 405-412.

Sugimoto, T., et al. 2016. Winter food habits of sympatric carnivores, Amur tigers and Far Eastern leopards, in the Russian Far East. - Mamm. Biol. 81: 214-218.

Sunquist, M. E. and Sunquist, F. 1989. Ecological constraints on predation by large felids. - In: Gittleman, J. L. (ed.) Carnivore behaviour, ecology and evolution. Cornell University Press, pp. 283-301.

Tian, Y., et al. 2011. Population viability of the Siberian Tiger in a changing landscape: Going, going and gone? - Ecol. Model. 222: 3166-3180.

Walston, J., et al. 2010. Bringing the tiger back from the brink-The six percent solution. - PLoS Biol. 8: e1000485.

Wang, T. M., et al. 2015. Long-distance dispersal of an Amur tiger indicates potential to restore the North-east China/Russian Tiger Landscape. - Oryx 49: 578-579.

Wang, T. M., et al. 2016. Amur tigers and leopards returning to China: direct evidence and a landscape conservation plan. - Landscape Ecol. 31: 491-503.

Wang, T. M., et al. 2016. Amur tigers and leopards returning to China: direct evidence and a landscape conservation plan. - Landscape Ecol. 31: 491-503.

Wang, T. M., et al. 2016. A science-based approach to guide Amur leopard recovery in China. - Biol. Conserv.: DOI: 10.1016/j.biocon.2016.1003.1014.

Wang, T. M., et al. 2018. Living on the edge: Opportunities for Amur tiger recovery in China. - Biol. Conserv. 217: 269-279.

Wang, T. M., et al. 2014. Camera traps reveal Amur tiger breeding in NE China. - Cat News. 61: 18-19.

Weckel, M., et al. 2006. Jaguar (*Panthera onca*) feeding ecology: distribution of predator and prey through time and space. - J. Zool. 270: 25-30.

Xiao, W. H., et al. 2016. Estimating abundance and density of Amur tigers along the Sino-Russian border. - Integr. Zool. 11: 322-332.

Xiao, W. H., et al. 2014. Distribution and abundance of Amur tiger, Amur leopard and their ungulate preys in Hunchun National Nature Reserve, Jilin. - Biodivers. Sci. 22: 717-724.

Zhang, J. L., et al. 2013. spaa: Species association analysis. R package version 0.2. 2. - <http://cran.r-project.org/package=spaa>.

Table 1. Prey species composition of tiger diets and their percent occurrence with 95% confidence interval, estimated weight ($X$), correction factor of weight per collected scat ($Y$) and biomass contribution in Hunchun Nature Reserve, Northeast China, between November 2014 and April 2015. Results are based on scat samples (n = 115) confirmed by genetic analysis from Amur tigers.

| Prey species | Percent occurrence (%) | $X$ (kg) | $Y$ (kg/scat) | Biomass contribution (%) |
|---|---|---|---|---|
| Wild boar | 52.16 (43.04-61.30) | 103 | 5.59 | 55.42 (46.60-64.29) |
| Sika deer | 26.29 (18.70-34.35) | 95 | 5.31 | 27.61 (20.22-36.23) |
| Roe deer | 13.36 (7.83-20.43) | 37 | 3.28 | 10.78 (6.30-16.74) |
| Badger | 2.59 (0.00-6.09) | 6 | 2.19 | 1.03 (0.33-2.85) |
| Cattle | 2.59 (0.00-6.09) | 331 | 13.57 | 2.88 (0.94-7.69) |
| Dog | 3.02 (0.87-6.96) | 31 | 3.07 | 2.27 (0.64-5.90) |

Table 2. List of tiger, leopard and potential prey species recorded by the camera traps, showing the number of trap stations, the number of independent detections and detection rate (number of detections per 100 trap nights) in Hunchun Nature Reserve, Northeast China.

| Species | Number of stations | % of all stations | Detections | % of all detections | Detection rate |
|---|---|---|---|---|---|
| Amur tiger | 42 | 40.38 | 131 | 8.82 | 0.77 |
| Amur leopard | 29 | 27.88 | 66 | 4.44 | 0.39 |
| **Wild ungulates** | | | | | |
| Wild boar | 47 | 45.19 | 106 | 7.14 | 0.62 |
| Roe deer | 54 | 51.92 | 177 | 11.92 | 1.04 |
| Sika deer | 45 | 43.27 | 210 | 14.14 | 1.23 |
| Musk deer | 3 | 2.88 | 19 | 1.28 | 0.11 |
| **Other wild prey** | | | | | |
| Black bear | 3 | 2.88 | 3 | 0.20 | 0.02 |
| Eurasian lynx | 1 | 0.96 | 1 | 0.07 | 0.01 |
| Leopard cat | 9 | 8.65 | 12 | 0.81 | 0.07 |
| Red fox | 39 | 37.50 | 137 | 9.23 | 0.8 |
| Raccoon dog | 9 | 8.65 | 25 | 1.68 | 0.15 |
| Badger | 24 | 23.08 | 62 | 4.18 | 0.36 |
| Siberian weasel | 10 | 9.62 | 21 | 1.41 | 0.12 |
| Sable | 1 | 0.96 | 3 | 0.20 | 0.02 |
| Yellow-throated marten | 15 | 14.42 | 26 | 1.75 | 0.15 |
| Hedgehog | 1 | 0.96 | 1 | 0.07 | 0.01 |
| Hare | 53 | 50.96 | 288 | 19.39 | 1.69 |
| **Domestic species** | | | | | |
| Cat | 1 | 0.96 | 1 | 0.07 | 0.01 |
| Dog | 44 | 42.31 | 172 | 11.58 | 1.01 |
| Cattle | 4 | 3.85 | 23 | 1.55 | 0.13 |
| Horse | 1 | 0.96 | 1 | 0.07 | 0.01 |

Table 3. Spatial and temporal overlap (95% confidence interval) between tigers and their main prey in Hunchun Nature Reserve, Northeast China.

| Variables | Spatial use | Temporal use | |
|---|---|---|---|
| | Pianka's index ($O$) | Overlap coefficient ($\Delta$) | Spearman's rho |
| Tiger and wild boar | 0.16 (0.08-0.30) | 0.76 (0.67-0.86) | 0.22* |
| Tiger and sika deer | 0.35 (0.23-0.54） | 0.69 (0.61-0.78) | -0.42* |
| Tiger and roe deer | 0.18 (0.05-0.45） | 0.74 (0.65-0.83) | 0.02 |

*$p < 0.01$

Table 4. The parameter estimates, standard error (SE), z value and p value from the occupancy model for the Amur tiger in Northeast China. Estimates of beta coefficients are reported for standardized covariates, scaled to mean and standard deviation. All of the reported estimates of coefficients that marked in bold are significant ($p < 0.05$).

| Covariate | Estimate | SE | z value | p value |
|---|---|---|---|---|
| Habitat use model | | | | |
| (Intercept) | 0.70 | 0.48 | 1.48 | 0.140 |
| Roe deer | **-0.72** | 0.33 | -2.17 | 0.030 |
| Sika deer | **1.99** | 0.83 | 2.40 | 0.017 |
| Wild boar | 0.18 | 0.31 | 0.57 | 0.568 |
| Detection model | | | | |
| (Intercept) | **-2.64** | 0.37 | -7.06 | <0.001 |
| Trail | **1.08** | 0.39 | 2.78 | 0.005 |
| Effort | -0.16 | 0.17 | -0.95 | 0.345 |

Table 5. Activity periods of the Amur tiger and its main prey based on the number of independent detections (N) recorded by the camera traps in Hunchun Nature Reserve, China, during November 2014-April 2015.

| Species | N | Diurnal (7:01-15:43, %) | Nocturnal (17:44-05:00, %) | Crepuscular (05:01-7:00 and 15:44-17:43, %) | Classification |
|---|---|---|---|---|---|
| Amur tiger | 131 | 21.2 | 48.5 | 30.3 | Cathemeral |
| Wild boar | 106 | 32.4 | 41.7 | 25.9 | Cathemeral |
| Sika deer | 210 | 49.8 | 27.2 | 23.0 | Mostly diurnal |
| Roe deer | 177 | 40.8 | 25.0 | 34.2 | Mostly diurnal |

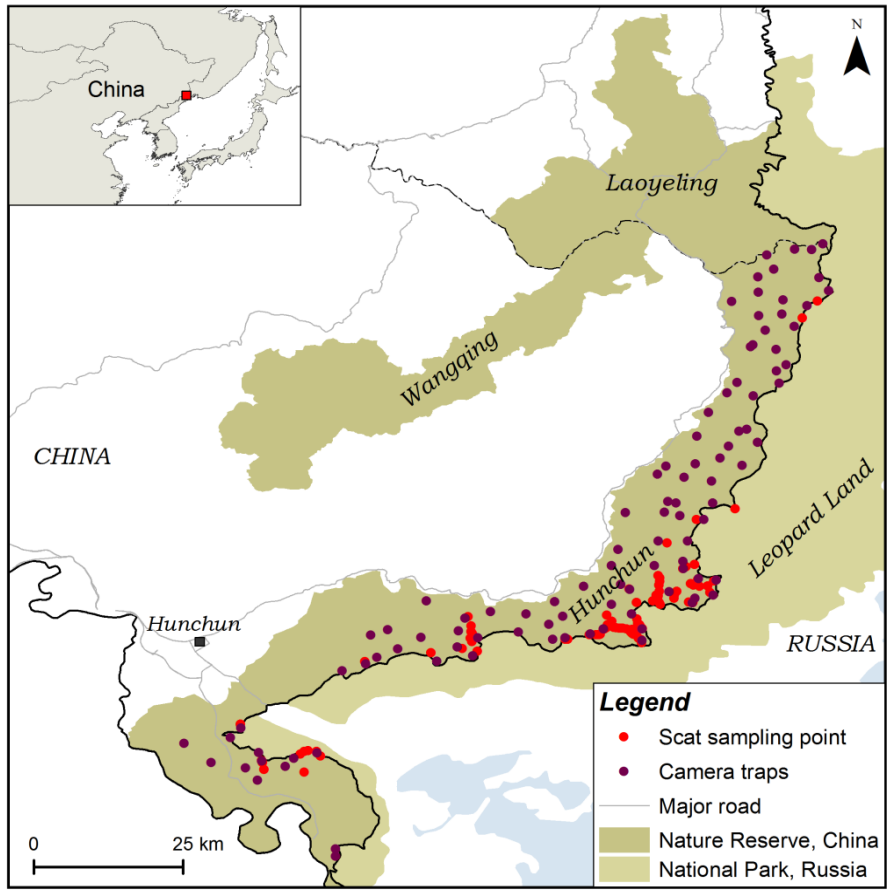

Figure 1. Map of the Hunchun Nature Reserve, Northeast China, with respect to scat sampling point, camera placement, main roads and adjacent national parks along the Sino-Russia border.

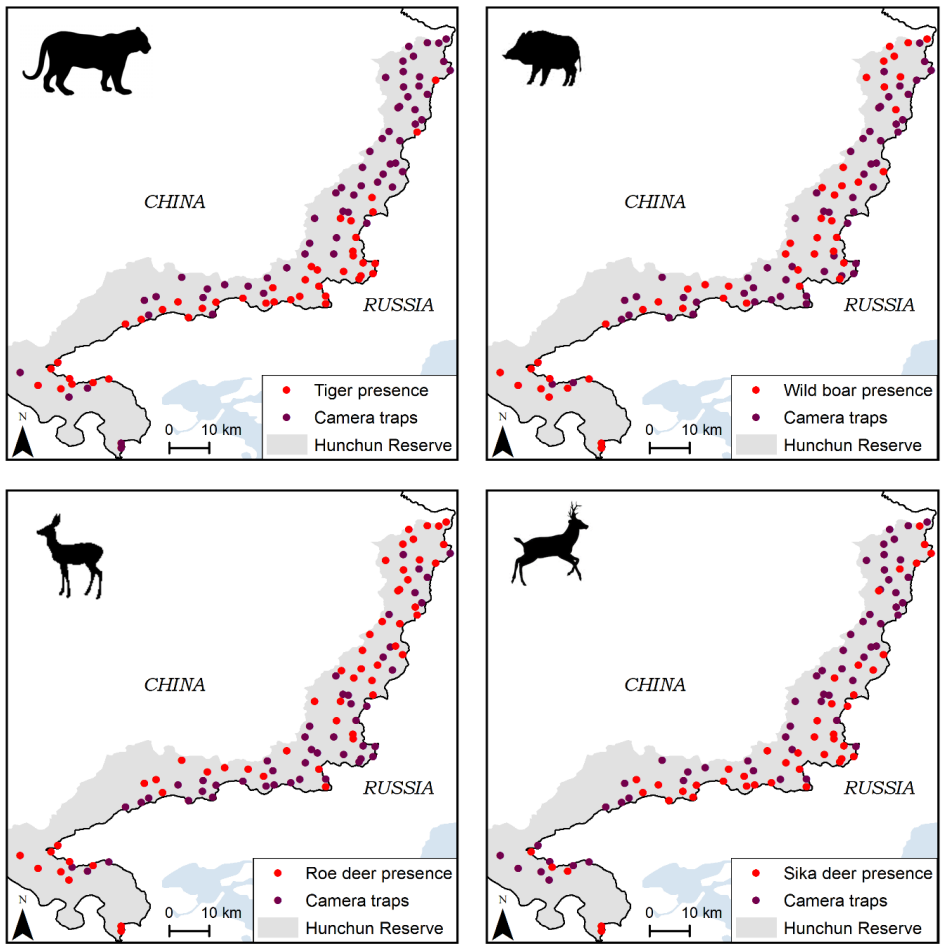

Figure 2. Spatial presence of tiger and three ungulate species in the Hunchun Natural Reserve, NE China. Purple dots represent sample locations (camera traps) where the species was not observed.

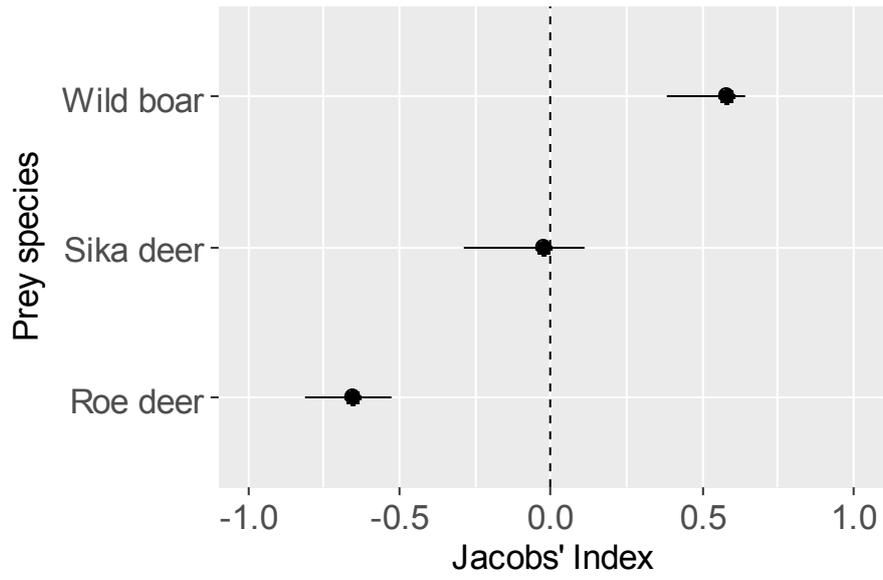

Figure. 3 Jacob's index (*D*) with 95% confidence interval shows tiger prey selection based on percent occurrence of prey species in Hunchun Nature Reserve, Northeast China.

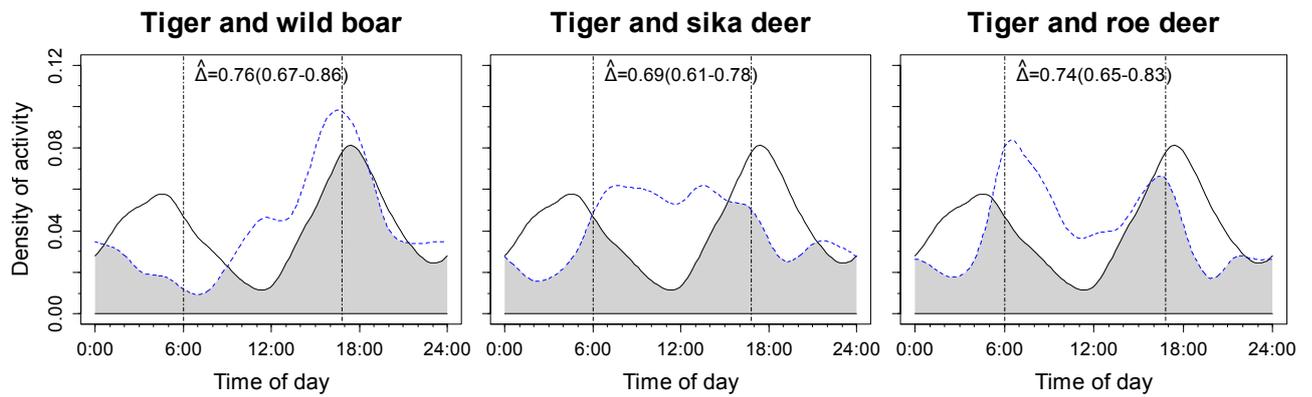

Figure 4. Temporal overlap of the daily activity patterns of tigers (solid lines) and their three main prey (blue dashed lines) in the Hunchun Nature Reserve of China. The estimated overlap coefficient ($\hat{\Delta}$) has values between 0 (no overlap) and 1 (complete overlap), with 95% bootstrap CIs in parentheses. Overlap is indicated by the shaded area. The grey dashed vertical lines indicate the approximate times of sunrise (6:00) and sunset (16:44) during the study period in study localities.

**Supplemental Information**

Table S1. Comparison of prey species percent biomass contribution to Amur tiger diets in Hunchun Nature Reserve (HNR) with three other Russian nature reserves, namely, Southwest Primorskii Krai (SW), Lazovsky State Nature Zapovednik (LZ) and Sikhote-Alin State Biosphere Zapovednik (SABZ), based on scat analysis (Kerley et al. 2015). HNR is bordered by SW to the east and LZ and SABZ are in the Sikhote-Alin mountain range.

| Species | HNR | SW | LZ | SABZ |
|---|---|---|---|---|
| Wild boar | 55.42 | 58.60 | 62.80 | 42.05 |
| Roe deer | 10.78 | 8.31 | 0.41 | 18.93 |
| Sika deer | 27.61 | 25.09 | 12.91 | 15.19 |
| Red deer | - | - | 6.47 | 13.72 |
| Musk deer | - | 0.37 | - | 0.76 |
| Long-tailed goral | - | - | 0.57 | - |
| Bear | - | 3.11 | 11.34 | 3.93 |
| Amur tiger | - | 2.81 | - | 2.66 |
| Raccoon dog | - | - | 1.12 | 0.95 |
| Badger | 1.03 | 0.34 | 2.55 | 1.15 |
| European otter | - | - | - | 0.67 |
| Siberian weasel | - | 0.94 | - | - |
| Cattle | 2.88 | - | - | - |
| Dog | 2.27 | 0.44 | 1.83 | - |